\documentclass[pra,aps,twocolumn,showpacs,superscriptaddress]{revtex4-1}

\usepackage{graphicx,amsmath,amsfonts,bbm,bm,braket,mathdots}
\usepackage{hyperref}
\usepackage[greek, english]{babel}

\usepackage[bordercolor=white,backgroundcolor=gray!30,linecolor=black,colorinlistoftodos]{todonotes}

\newcommand{\gpi}{\text{\greektext p}}	
\newcommand{\ii}{\mathrm{i}}	        
\newcommand{\ee}{\mathrm{e}}    	    
\newcommand{\dd}{\mathrm{d}}		    
\renewcommand{\Re}{\mathrm{Re}\,}       
\renewcommand{\Im}{\mathrm{Im}\,}       
\renewcommand{\vec}[1]{\boldsymbol{#1}}	
\renewcommand{\t}[1]{\text{#1}}	        
\newcommand{\identity}{\mathbbm{1}}     
\newcommand\varpm{\mathbin{\vcenter{\hbox{\oalign{\hfil$\scriptscriptstyle+$\hfil\cr\noalign{\kern-.3ex}$\scriptscriptstyle({-})$\cr}}}}}
\newcommand\varmp{\mathbin{\vcenter{\hbox{\oalign{$\scriptscriptstyle({+})$\cr\noalign{\kern-.3ex}\hfil$\scriptscriptstyle-$\hfil\cr}}}}}

\begin{document}

\title{Topological invariants in dissipative extensions of the Su-Schrieffer-Heeger model}

\author{Felix Dangel}
\email{felix.dangel@itp1.uni-stuttgart.de}
\thanks{}
\affiliation{Institut f\"ur Theoretische Physik 1, Universit\"at Stuttgart,
	70550 Stuttgart, Germany}
\author{Marcel Wagner}
\email{marcel.wagner@itp1.uni-stuttgart.de}
\thanks{\\${}^{\ast \dagger}$\ both authors contributed equally to this work}
\affiliation{Institut f\"ur Theoretische Physik 1, Universit\"at Stuttgart,
	70550 Stuttgart, Germany}
\author{Holger Cartarius}
\affiliation{Institut f\"ur Theoretische Physik 1, Universit\"at Stuttgart,
	70550 Stuttgart, Germany}
\affiliation{Physik und ihre Didaktik, Universit\"at Stuttgart,
	70550 Stuttgart, Germany}
\author{J\"org Main}
\affiliation{Institut f\"ur Theoretische Physik 1, Universit\"at Stuttgart,
	70550 Stuttgart, Germany}
\author{G\"unter Wunner}
\affiliation{Institut f\"ur Theoretische Physik 1, Universit\"at Stuttgart,
	70550 Stuttgart, Germany}

\date{\today}

\begin{abstract}
    We investigate dissipative extensions of the Su-Schrieffer-Heeger model with regard to different approaches of modeling dissipation. 
In doing so, we use two distinct frameworks to describe the gain and loss of particles, one uses Lindblad operators within the scope of Lindblad master equations, the other uses complex potentials as an effective description of dissipation.
The reservoirs are chosen in such a way that the non-Hermitian complex potentials are $\mathcal{PT}$-symmetric.
From the effective theory we extract a state which has similar properties as the non-equilibrium steady state following from Lindblad master equations with respect to lattice site occupation. 
We find considerable similarities in the spectra of the effective Hamiltonian and the corresponding Liouvillean. 
Further, we generalize the concept of the Zak phase to the dissipative scenario in terms of the Lindblad description and relate it to the topological phases of the underlying Hermitian Hamiltonian. \end{abstract}

\maketitle

\section{Introduction}

The question of how dissipation influences topological quantum systems is heavily addressed in today's research \cite{Esaki2011ZakPhaseNonHermitian,Bardyn2013a,Diehl2011,Budich2015DissipativePreparation,Weimann2017EdgeStatesPhotonicCrystal}. 
In literature various concepts are used and proposed to generalize the cherished tools developed for a characterization of topological phases in closed quantum systems to open systems.
In this field two different approaches of describing dissipation are frequently used. 
One is based on Lindblad operators, which describe the interactions of a system with a reservoir, and allow for the calculation of the time evolution of the system's density matrix via \emph{Lindblad master equations} (LME) \cite{Lindblad1976}.
This framework has been applied to prepare quantum systems in topologically nontrivial states by engineering the dissipative dynamics \cite{Budich2015DissipativePreparation,Diehl2011}. 
Also generalizations of topological invariants have been formulated in the course of this framework and the effects of dissipation on the topological properties of open quantum systems have been investigated \cite{Bardyn2013a,Rivas2013DensityMatrixChernInsulators,Budich2015DensityMatrices,Linzner2016ThoulessPumping,Grust2017TopologicalOrder}.

Another approach of describing dissipation uses complex potentials and effective non-Hermitian Hamiltonians $H_\text{eff}$ to model the gain and loss of particles.
In this context reservoirs which are invariant under a spatial inversion $\mathcal{P}$ and a simultaneous time reflection $\mathcal{T}$ (interchange of particle sinks and sources) are of special interest. Such systems can be effectively described by $\mathcal{PT}$-symmetric Hamiltonians, which then fulfill $\left[ H_\text{eff}, \mathcal{PT} \right]  = 0$ and, in spite of their non-Hermiticity, still can possess entirely real eigenvalue spectra \cite{BenderBoettcher1998RealSpectra}.
Within this description the effects of dissipation on topological systems have been investigated and it has been debated, whether or not topologically nontrivial states may exist in combination with $\mathcal{PT}$-symmetric gain-loss patterns \cite{Hu2011AbsenceToplogicalPhases,Esaki2011ZakPhaseNonHermitian,Ghosh2012TopologicalPhaseNonHermitian,Zhu2014PTNonHermitianSSH,Yuce2015PTFloquetTopological,Klett2017PTSymmetry,Menke2017TopologicalQuantumWires}. Special interest is triggered in optics \cite{Schomerus2013PhotonicLattices}, where experimental realizations have successfully been developed \cite{Zeuner2015ObservationBulk,Weimann2017EdgeStatesPhotonicCrystal}. While most of the works address the appearance of edge states, topological invariants have also been formulated for the non-Hermitian case \cite{Esaki2011ZakPhaseNonHermitian,Menke2017TopologicalQuantumWires,we2017,Lieu2018}.

The generic example of a one-dimensional topological insulator is the \emph{Su-Schrieffer-Heeger} (SSH) model \cite{SSH1979} which was initially proposed to study solitons in polyacetylene.
In the present paper we compare both approaches by way of the example of the SSH model subject to gain and loss. A first comparison was done with regard to the appearance of edge states \cite{Klett2018MasterPT}. Here we want to go one step further and study whether the characterization of topological phases in both approaches leads to an agreement. To do so, we introduce two different $\mathcal{PT}$-symmetric dissipative patterns and investigate a topological invariant, viz.\ the real part of the complex Zak phase. 
We find that the topological invariants of both approaches coincide in the parameter regime where the effective theory  possesses an unambiguous interpretation.
Further we find a remarkable accordance in the long-term dynamics following from both approaches, where we justify a construction scheme of a fixed-point-like many-body state in the effective $\mathcal{PT}$-symmetric theory. 

Since we combine disparate descriptions and methods, the first part of the paper (Secs.\ \ref{sec:ssh} - \ref{sec:ComplexBerry}) is dedicated to give an overview of the methods and concepts used in the analysis.
We shortly summarize the most important aspects of the SSH model in Sec.\ \ref{sec:ssh} before we introduce the dissipative frameworks in Sec.\ \ref{sec:DissipativeFrameworks}. 
For spatially periodic systems (periodic boundary conditions) we introduce a momentum basis, which allows for the generalization of the Zak phase to dissipative systems described by an LME. 
The method of expressing a Liouvillean in a momentum basis was previously used in \cite{Rivas2013DensityMatrixChernInsulators} to generalize the Chern number to dissipative systems. 
In Sec.\ \ref{sec:ComplexBerry} we argue that in case of the effective description of dissipation the real part of the complex Zak phase is quantized and can be used as a topological invariant, the corresponding topological phases of which are protected by $\mathcal{PT}$ symmetry.
The results of the comparison of the descriptions are presented in Sec.\ \ref{sec:Comparison}, where we find clear similarities in both approaches for modeling dissipation. \section{\label{sec:ssh}SSH model}

\begin{figure}
	\centering
		\includegraphics[page = 1, width = \linewidth]{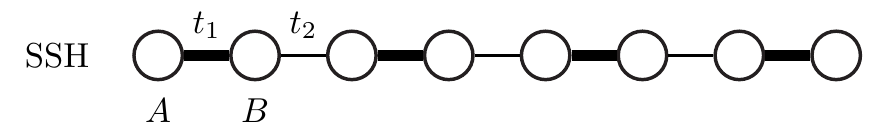}
		\vspace{3ex}
		\includegraphics[page = 3, width = \linewidth]{Fig1.pdf}
		\vspace{3ex}
		\includegraphics[page = 2, width = \linewidth]{Fig1.pdf}
		\vspace{-3ex}
	\caption{Sketch of the SSH model and its extensions for $n=8$ sites. Top row: The Hermitian model consists of double-well unit cells (sites $A, B$) with intra-tunneling $t_1$ joined by the inter-cell hopping $t_2$. Lower side: In this work we study dissipative extensions of the SSH model where sites marked with a plus (minus) sign indicate single-particle gain (loss). The two patterns with dissipation only among the boundary sites ($U_1$, middle row) and alternating gain and loss ($U_2$, bottom row) are motivated by previous works \cite{Zhu2014PTNonHermitianSSH, Wang2015SpontaneousPTBreaking, Menke2017TopologicalQuantumWires, Weimann2017EdgeStatesPhotonicCrystal, Klett2017PTSymmetry}.}
	\label{fig:model}
\end{figure}

The paradigmatic one-dimensional SSH model \cite{SSH1979} describes spin-polarized non-interacting fermions on a lattice with $n$ sites arranged in double-well unit cells in such a way that nearest-neighbor tunneling alternates between $t_1$ and $t_2$ (see Fig.\ \ref{fig:model}). For comparison with the dissipative extensions considered in this work its properties are briefly outlined. The Hamiltonian $H$ describing the SSH model is given by
\begin{align}\label{eq:HermSSH}
	\begin{split}
		H & =-\sum_{j=1}^{n/2} \left( t_1 c_{2j-1}^{\dagger} c_{2j}^{\phantom{\dagger}} +  t_2 c_{2j}^{\dagger} c_{2j+1}^{\phantom{\dagger}} + \t{h.c.}\right)
		\\
		& = -\sum_{j=1}^{n/2} \Big( t_1  \ket{j,A} \! \bra{j,B} 
		 + t_2 \ket{j,B} \! \bra{j+1,A} + \t{\t{h.c.}}\Big),
	\end{split}\hspace*{-0.35cm}
\end{align}
where $c_i^{\phantom{\dagger}}$ $(c_{i}^{\dagger})$ denotes the annihilation (creation) operator of a spinless fermion at the site $i$. 
For staggered hopping amplitudes $t_1 \neq t_2$, the Hamiltonian in Eq.\ \eqref{eq:HermSSH} describes a two-band topological insulator \cite{HassanKane2010TopologicalInsulators} that undergoes a topological phase transition at $t_1 = t_2$ where the band gap closes. 
In fact, the periodic translation invariant system can be solved analytically by Fourier transformation $\ket{k,X} = 1/\sqrt{n/2} \sum_j \ee^{\ii j k} \ket{j,X}$ of the unit cell index $j$, where $X \in \{A, B\}$ and $k = 0, 2\gpi/(n/2), 4\gpi/(n/2), \dots, 2\gpi (n/2-1)/(n/2)$. 
Fac\-torizing $\ket{k,X} = \ket{k} \otimes \ket{X}$, the Hamiltonian is fully described by the $2\times 2$ \emph{Bloch Hamiltonian} $H_\text{B}$ (the matrix occurring on the right side),
\begin{equation}
	\label{equ:sshBlochHamiltonian}
	H = \sum_k \ket{k} \! \bra{k} \otimes 
	\begin{pmatrix}
		0 & -t_1 -t_2\ee^{\ii k}
		\\
		-t_1 - t_2 \ee^{-\ii k} & 0
	\end{pmatrix} ,
\end{equation}
where the sum runs over the discrete steps of $k$ mentioned above.
The Bloch Hamiltonian yields the energies $E_m(k) = \pm \sqrt{t_1^2 + t_2^2 + 2 t_1 t_2 \cos(k)}$. 

The topological invariant of the Bloch band corresponding to $\ket{u_m(k)}$ (an eigenvector of the Bloch Hamiltonian) is given by the \emph{Berry phase} $\nu_m$ \cite{Berry1984BerryPhase} in momentum space, also known as \emph{Zak phase} \cite{Zak1989ZakPhase}
\begin{equation}
	\nu_m = \ii \oint_{0}^{2\pi} \!\!\!\!\!\! \braket{u_m(k) | \partial_k | u_m(k)} \dd k
\end{equation}
and takes a value of $\nu_m = \gpi$  if $t_1 < t_2$ and $\nu_m = 0$ if $t_1 > t_2$ \cite{Asboth2016TopologicalInsulators}. In the topologically \emph{nontrivial} phase ($\nu_m \neq 0$) the SSH model possesses zero-energy edge modes at open boundaries, which are protected by the topological properties of the system.

 \section{\label{sec:DissipativeFrameworks}Dissipative frameworks}

In the further course of this paper we allow particles to enter respectively exit the system on certain sites as sketched in the lower part of Fig.\ \ref{fig:model}. This notion of dissipation is modeled by two different approaches.

\subsection{Lindblad master equation}

The LME \cite{Lindblad1976} results from the Markovian approximation \cite{Breuer2007OpenQuantumSystems} of the reservoir and describes the dissipative (trace and positivity preserving) evolution of the density matrix $\rho$,
\begin{equation}\label{eq:LindbladMaster}
	\dot{\rho} = -\ii \big[ H, \rho \big] + \sum_{\mu} \big(2 L_{\mu}^{\phantom{\dagger}} \rho L_{\mu}^{\dagger} - \big\{ L_{\mu}^{\dagger}L_{\mu}^{\phantom{\dagger}}, \rho \big\} \big) \equiv \hat{\mathcal{L}} \rho,
\end{equation}
where the unitary evolution generated by $H$ is supplemented by the influence of collapse operators $L_\mu$ characterizing the dissipative coupling to the reservoir, and we have set \mbox{$\hbar \equiv 1$.}
Writing Eq.\ \eqref{eq:LindbladMaster} as an operator equation introduces the Liouville operator (or Liouvillean) $\hat{\mathcal{L}}$. In the long-time limit, the system converges to the \emph{non-equilibrium steady state} (NESS), which satisfies $\hat{\mathcal{L}} \rho_\text{ness} = 0$.

In this work, we choose the Lindblad operators
$L_{\mu}=\sqrt{\gamma} c_\mu^{\dagger}$ $(\sqrt{\gamma} c_\mu^{\phantom{\dagger}})$ to describe single-particle gain (loss) with a rate $\gamma$.
Consequently, the dissipative patterns presented in Fig.\ \ref{fig:model} are expressed by the following choice of Lindblad couplings ($\mu = 1, \dots, n$),
\begin{subequations}
	\begin{align}
	\label{eq:LindbladOpsU1}
	\hspace*{-2mm} U_1: \quad \hspace*{0.3mm} L_1 &= \sqrt{\gamma} c_{1}^{\phantom{\dagger}}, \quad
	L_n= \sqrt{\gamma} c_{n}^{\dagger}, \quad L_{\mu}=0 \ \text{(else)} ,
	\\
	\label{eq:LindbladOpsU2}
	\hspace*{-2mm} U_2: \quad L_\mu &= \sqrt{\gamma} c_{\mu}^{\dagger} \ \text{($\mu$ odd)}, \quad
	L_\mu = \sqrt{\gamma} c_{\mu}^{\phantom{\dagger}} \ \text{($\mu$ even)}.
	\end{align}
\end{subequations}

We note that the reservoir $U_2$ does not spoil the translational symmetry and the appropriate system can still be naturally described in momentum space. \subsubsection*{\label{sec:3rdQZ}Third quantization}

Any fermionic dissipative system described by a master equation in Lindblad form \eqref{eq:LindbladMaster} with a quadratic Hamiltonian $H$ and linear collapse operators $L_\mu$ can be treated by means of a method named \emph{third quantization}, presented in references \cite{Prosen2008ThirdQuantizationFermions, Prosen2008XYChain}: 
Both constituents of the Liouvillean $\hat{\mathcal{L}}$ are expanded in terms of $2n$ \emph{abstract Hermitian Majorana operators} $w_j$ (which are in our case related to the fermionic operators by $w_{2m-1} = c_m + c_m^\dagger, w_{2m} = \ii(c_m - c_m^\dagger)$ with $m = 1, \dots, n$) as $H = \sum_{j,k = 1}^{2n} w_j H_{jk} w_k$ and $L_\mu = \sum_{j = 1}^{2n}l_{\mu, j} w_j$. 
The operator space $\mathcal{K}$ of the $w_j$ is spanned by the $2^{2n}$-dimensional orthonormal basis vectors $P_{\alpha_1,\dots,\alpha_{2n}} = w_1^{\alpha_1} \cdots w_{2n}^{\alpha_{2n}}$ with $\alpha_j \in \{0, 1\}$.
By introducing super-operators on $\mathcal{K}$ in terms of \emph{adjoint Fermi maps} $\hat{c}_j, \hat{c}_j^\dagger\ (j = 1, \dots, 2n)$ that act on the canonical basis as $\hat{c}_j \ket{P_{\alpha_1,\dots,\alpha_2n}} = \delta_{\alpha_j,1} \ket{w_j P_{\alpha_1,\dots,\alpha_2n}}, \hat{c}_j^\dagger \ket{P_{\alpha_1,\dots,\alpha_2n}} = \delta_{\alpha_j,0} \ket{w_j P_{\alpha_1,\dots,\alpha_2n}}$, the Liouvillean can be rewritten and becomes bilinear after a linear transformation to $4n$ \emph{adjoint Hermitian Majorana maps} $\hat{a}_{2j-1} = (\hat{c}_j + \hat{c}_j^\dagger)/\sqrt{2}$, $\hat{a}_{2j} = \ii (\hat{c}_j - \hat{c}_j^\dagger)/\sqrt{2}$, where $j = 1 \dots, 2n$. 
The resul\-ting expression,
\begin{equation}
	\hat{\mathcal{L}} = \sum_{i,j=1}^{4n} \hat{a}_i A_{ij} \hat{a}_j - A_0 \hat{\identity},
\end{equation}
introduces the antisymmetric $4n\times 4n$ \emph{shape matrix} $\bm{A} = - \bm{A}^T$, which contains all information about the system. 
Its eigenvalues, referred to as \emph{rapidities}, come in pairs $\pm\beta_{1,\dots, 2n}$ with $\mathrm{Re}(\beta_j) \ge 0$. 
By means of the shape matrices' eigenvectors, the dissipative system decomposes into \emph{normal master modes} (NMM) that are populated at an exponential rate given by $2\Re(\beta_j)$ \cite{Prosen2008ThirdQuantizationFermions}. 
Hence, the NESS is unique whenever all rapidities $\beta_j$ satisfy $\Re(\beta_j) > 0$. As shown in reference \cite{Prosen2008ThirdQuantizationFermions}, NESS expectation values can be computed straightforwardly in this case.

Applying this procedure to an SSH ring with Lindblad operators $\sqrt{\gamma} c_\mu, \sqrt{\gamma}c_\mu^{\dagger}$ arranged in the alternating pattern of $U_2$, the shape matrix takes the banded form 
\begin{subequations} \label{eq:ShapePeriodicBC}
	\begin{align}
	\bm{A} &= \frac{1}{2}\begin{pmatrix}
					\gamma \bm{\Gamma}_\text{g} & -t_1 \bm{T}& & & & -t_2 \bm{T}
					\\
					-t_1 \bm{T} & \gamma \bm{\Gamma}_\text{l} & -t_2 \bm{T} & 
					\\
					& -t_2 \bm{T} & \gamma \bm{\Gamma}_\text{g} & -t_1 \bm{T} & 
					\\
					&  & \ddots & \ddots & \ddots
					\\
					-t_2 \bm{T}
		\end{pmatrix},
	\end{align}
	with $4\times 4$ matrices $\bm{\Gamma}_\text{g(l)} = - \identity_{2} \otimes \sigma_y \varpm \sigma_y \otimes \left( \ii \sigma_x  + \sigma_z\right)$ and  $\bm{T} = -\ii\sigma_y \otimes \identity_{2}$. 
	As the dissipative pattern does not spoil the system's unit cell structure, $\bm{A}$ is naturally expressed by partitioning the matrix into $8\times 8$ blocks labeled with $j = 1, \dots, n/2$, which themselves consist of $4\times4$ blocks $A,B$. 
	Adopting a projector notation the shape matrix reads
		\begin{align}
			\begin{split}
				\bm{A} = \frac{1}{2}\sum_{j=1}^{n/2} &\Big( \gamma \big[ \ket{j, A}\bra{j, A} \otimes \bm{\Gamma}_\text{g}
				+ \ket{j, B}\bra{j, B} \otimes \bm{\Gamma}_\text{l}\big]
				\\
				&-t_1 \big[ \ket{j,A}\bra{j,B} + {\t{h.c.}} \big] \otimes \bm{T}
				\\
				&-t_2 \big[ \ket{j,B}\!\bra{j+1,A} + {\t{h.c.}} \big] \otimes  \bm{T}\Big),
			\end{split} 
		\end{align}
	which resembles the form of the $\mathcal{PT}$-symmetric Hamiltonian subject to $U_2$ mentioned below with matrices $\bm{T}, \bm{\Gamma}_\text{g}, \bm{\Gamma}_\text{l}$ replacing the scalar entries of tunneling amplitudes and gain and loss rates, respectively. 
	In analogy with the Hamiltonian case, the representation can be further compacted by transforming the external degree of freedom into momentum space with a Fourier transform, $\ket{k, X} = 1/\sqrt{(n/2)} \sum_{j = 1}^{n/2} \mathrm{e}^{\ii j k} \ket{j, X}$ with $k = 0, 2\gpi/(n/2), 4\gpi/(n/2), \dots, 2\gpi (n/2-1)/(n/2)$ and $X \in \{A, B\}$, resulting in	
		\begin{align}
			\label{eq:BlochLiouvillean}
			\begin{split}
				\bm{A} = \frac{1}{2} \sum_k \ket{k}\!\bra{k} \otimes& \Big(\gamma \ket{A}\!\bra{A} \otimes \bm{\Gamma}_\text{g} + \gamma \ket{B}\!\bra{B} \otimes \bm{\Gamma}_\text{l}
				\\
				&- \big[(t_1 + \mathrm{e}^{\ii k} t_2) \ket{A}\!\bra{B}  + \text{h.c.}\big] \otimes \bm{T}\Big)
				\\
				= \frac{1}{2} \sum_k \ket{k}\!\bra{k} \otimes& \begin{pmatrix}
				 \gamma \bm{\Gamma}_\text{g} & -(t_1 + t_2 \mathrm{e}^{\ii k}) \bm{T}
				 \\
				 -(t_1 +  t_2 \mathrm{e}^{-\ii k}) \bm{T} & \gamma \bm{\Gamma}_\text{l} 
				\end{pmatrix},
			\end{split}
		\end{align}	
\end{subequations}
where the additional factorization $\ket{k, X} = \ket{k} \otimes \ket{X}$ has been assumed.
Due to the similarity of this procedure and the derivation of the Bloch Hamiltonian, we will refer to the matrix in the last equation as \emph{Bloch Liouvillean}. \subsection{\label{subsec:effectiveTheory}\texorpdfstring{\bm{$\mathcal{PT}$}}{PT}-symmetric potentials (effective theory)}

Our second approach is given by a description of dissipation using complex on-site potentials that lead to an effective non-Hermitian Hamiltonian and in fact triggered the surge of the entire research area of non-Hermitian quantum mechanics \cite{Moiseyev2011NonHermitianQM, Brody2014BiorthogonalQuantumMechanics}. 
This procedure has already been promisingly applied to bosonic systems \cite{Dast2014BalancedGainAndLoss}. 
It can be motivated by considering a single site with $H=0$ subject to single-particle gain (loss) described by an LME \eqref{eq:LindbladMaster}, which reads $\dot{\rho}= \gamma (2 c^{\dagger} \rho c - c c^{\dagger} \rho - \rho c c^{\dagger})$ (with $c \leftrightarrow c^\dagger$ swapped for single-particle loss). 
The populations $\rho_{ii}, i = 0, 1$ of the solution $\rho(t) = \sum_{i,j} \rho_{ij}(t) \ket{i}\!\bra{j}$  show an exponential decrease at the rate of $2\gamma$, that is $\rho_{11}(t) = \rho_{11}(0)\ee^{-2\gamma t}$ in the case of single-particle loss, and analogously for the gain scenario.
However, such an exponential increase (except for the maximum occupation) can also be implemented by a complex on-site potential $\pm \ii \gamma c^\dagger c$, which becomes exact in the mean-field limit of bosons \cite{Dast2014BalancedGainAndLoss}. 

Hence, we account for single-particle gain (loss) at site $j$ via the extension of the Hamiltonian with a term $\varpm ̣̣̣̣\ii \gamma c_j^\dagger c_j$ to obtain an effective description of dissipation. 
This results in a non-Hermitian Hamiltonian $H_{\text{eff}}^{(U)} = H + U$. The complex potentials, which effectively describe the effects of the reservoirs shown in Fig.\ \ref{fig:model} are given by
\begin{subequations}
	\label{equ:PotentialsPT}
	\begin{align}
		\label{eq:PotU1}
		U_{1} &= \ii \gamma \left( c_{n}^{\dagger}c_{n}^{\phantom{\dagger}} - c_{1}^{\dagger}c_{1}^{\phantom{\dagger}}\right),
		\\
		\label{eq:PotU2}
		U_{2} &=\sum \limits_{j=1}^{n/2} \ii \gamma \left( c_{2j-1}^{\dagger}c_{2j-1}^{\phantom{\dagger}} - c_{2j}^{\dagger}c_{2j}^{\phantom{\dagger}}\right).
	\end{align}
\end{subequations}
A helpful property of both reservoirs in Eq.\ \eqref{equ:PotentialsPT} as well as the Hamiltonian in Eq.\ \eqref{eq:HermSSH} is their invariance under the combined action of parity and time inversion, $\left[ H_\text{eff}, \mathcal{PT} \right]  = 0$, which causes the complex energy spectrum to be entirely real-valued in the $\mathcal{PT}$-unbroken parameter regime.

A further analogy between the two approaches can be revealed by following the physical interpretation of a prominent algorithm, which allows for the computation of the time evolution of observables of systems characterized by an LME. 
The so called \emph{Monte Carlo wave-function} approach uses the combination of a time evolution with a non-Hermitian Hamiltonian and quantum jumps to determine steady states of an open quantum system described by an LME \cite{Molmer1993QuantumMonteCarlo}. 
For $\mathcal{PT}$-symmetric systems the non-Hermitian Hamiltonian constructed within the numerical method is equivalent to the $\mathcal{PT}$-symmetric Hamiltonian of the effective approach in Sec.\ \ref{subsec:effectiveTheory}, except for a constant imaginary shift. Thus, our effective $\mathcal{PT}$-symmetric Hamiltonian is connected with the non-Hermitian Hamiltonian of the Monte Carlo wave-function algorithm. \section{\label{sec:ComplexBerry}Complex Berry phase}

In the scope of the effective $\mathcal{PT}$-symmetric theory the concept of the Berry phase can be generalized to dissipative systems \cite{Garrison1988, Nenciu1992, Mostafazadeh1999AdiabaticCyclicStatesNonHermitian, Esaki2011ZakPhaseNonHermitian}.
In case of spatially periodic systems like $U_2$ the \emph{complex Zak phase} can be defined with the help of pairs of biorthogonal eigenvectors of the non-Hermitian $\mathcal{PT}$-symmetric Hamiltonian $H_\text{eff}^{(U_2)}$ belonging to real eigenvalues. To do so, one follows the same procedure as described in Sec.\ \ref{sec:ssh}, which results in
\begin{align}\label{eq:SSHU2Bloch}
	H_{\text{eff}}^{(U_2)} & = \sum_{k}^{} \ket{k}\!\bra{k} \otimes
	\begin{pmatrix}
		\ii \gamma & -t_1-t_2 \ee^{\ii k}
		\\
		-t_1-t_2 \ee^{-\ii k} & -\ii \gamma
	\end{pmatrix},
\end{align}
and energies $E_m(k) = \pm \sqrt{t_1^2 + t_2^2 + 2t_1t_2\cos(k) - \gamma^2}$, which are entirely real as long as $|t_1 - t_2| > \gamma$ \cite{Weimann2017EdgeStatesPhotonicCrystal}. 
The $2\times 2$ matrix represents the $\mathcal{PT}$-symmetric Bloch Hamiltonian. 
The structure of the Bloch Liouvillean found in Eq.\ \eqref{eq:BlochLiouvillean} is very similar to the one of the non-Hermitian Bloch Hamiltonian in Eq.\ \eqref{eq:SSHU2Bloch}.
In case of unbroken $\mathcal{PT}$ symmetry, that is if all eigenvalues of $H_{\text{eff}}$ are real-valued, the complex Zak phase which is picked up by the $m$th Bloch band, described by the left and right eigenvectors $\bra{\chi_m}$ and $\ket{\phi_m}$ of the Bloch Hamiltonian associated with eigenvalue $E_m$, is defined as \cite{Garrison1988, Mostafazadeh1999AdiabaticCyclicStatesNonHermitian}
\begin{equation} 
	\label{eq:complexBerry}
	\nu_{m}= \ii \oint_{0}^{2\gpi} \!\!\!\!\!\! \bra{\chi_m(k)} \partial_k \ket{\phi_m(k)} \dd k.
\end{equation} 
In the $\mathcal{PT}$-unbroken regime, the real part of the complex Zak phase is quantized, $\Re(\nu_m) = 0, \gpi \mod2\gpi$, as shown in App.\ \ref{app:quanizationBerry}. 
In analogy with the line of argument of Hatsugai \cite{Hatsugai2006} for Hermitian systems we use the quantized real part of the complex Zak phase as topological invariant to characterize topological phases in the periodic lattice system described by Eq.\ \eqref{eq:SSHU2Bloch}. 
The quantization of the real part of the complex Zak phase is ensured by $\mathcal{PT}$ symmetry, and thus the corresponding topological phases are protected by $\mathcal{PT}$ symmetry (see App.\ \ref{app:quanizationBerry}).
  
In addition to analytical results of the Berry phase defined by Eq.\ \eqref{eq:complexBerry} in Ref.\ \cite{Esaki2011ZakPhaseNonHermitian}, an algorithm to numerically evaluate the expression is described in Ref. \cite{we2017}. 
Note, however, that the extension of Berry phases is limited to the $\mathcal{PT}$-unbroken regime and if the eigenvalues of $H_\text{eff}$ are complex, the adiabatic theorem which is used in the derivation of the complex Zak phase does not apply \cite{Nenciu1992} and a Berry phase is not well-defined. 

Note that the notion of a chiral symmetry $\Lambda = \sigma_z$ protecting the topology in the Hermitian case $\gamma = 0$ where $\left\{H_\text{B}, \sigma_z \right\} = 0$ cannot be directly carried over to the dissipative case due to the non-Hermiticity of the Hamiltonian. In fact, the underlying relations have to be modified for non-Hermitian Hamiltonians \cite{Martinez2018Symmetries} and it is nevertheless possible to find a symmetry operator that constrains the eigenvalue spectrum of $H_\text{eff}^{(U_2)}$ in the same way as the chiral symmetry does in the Hermitian SSH model.

To see this, consider the Hamiltonian of Eq. \eqref{eq:SSHU2Bloch} with open boundaries expanded in the $n$-dimensional single-particle basis,
	\begin{align}
		H_\text{eff}^{(U_2)} = 	\begin{pmatrix}
									\ii \gamma & - t_1 & 0 
									\\
									- t_1 & -\ii \gamma & -t_2 & \ddots
									\\
									0 & -t_2 & \ii \gamma & \ddots
									\\
									& \ddots & \ddots & \ddots 
								\end{pmatrix}.
	\end{align}
Using the unitary $n$-dimensional operators
\begin{align}
	\Sigma_x = 	\begin{pmatrix}
					& & 1
					\\
					& 1 &
					\\
					\iddots & &
				\end{pmatrix}, 
				\qquad 
					\Sigma_z =  \begin{pmatrix}
					1 & & 
					\\
					& -1 & 
					\\
					& & \ddots					
				\end{pmatrix}
\end{align}
introduced in the Suppl.\ Mat.\ of Ref.\ \cite{Weimann2017EdgeStatesPhotonicCrystal} one can construct the \emph{non-Hermitian} operator $\Lambda = \Sigma_x \Sigma_z = -\Lambda^\dagger$ that satisfies the relation
\begin{align}
	\label{eq:SymmetryProperty}
	\Lambda^\dagger H_\text{eff}^{(U_2)} \Lambda =  - H_\text{eff}^{(U_2)}.
\end{align}
Similarly to the chiral symmetry in the Hermitian scenario the symmetry property \eqref{eq:SymmetryProperty} constrains the eigenvalue spectrum of $H_\text{eff}^{(U_2)}$ to bands of opposite sign: Given left and right eigenvectors $\bra{\chi}, \ket{\phi}$ of $H_\text{eff}^{(U_2)}$ associated with an eigenvalue $E$, the symmetric partner states $\bra{\chi}\Lambda^\dagger, \Lambda \ket{\phi}$ are eigenvectors with an eigenvalue of $-E$.
By the same reasoning this property carries over to the spectrum of $H_\text{eff}^{(U_1)}$.

As a side mark, note that it is possible to construct non-Hermitian $\mathcal{PT}$-symmetric systems with alternating gain and loss and a centered defect that hosts topologically protected zero-energy edge states in the $\mathcal{PT}$-unbroken phase \cite{Weimann2017EdgeStatesPhotonicCrystal}. Their stability under disorder respecting the symmetry $\Lambda = \Sigma_x \Sigma_z$ has been verified analytically and numerically in the Suppl.\ Mat.\ of Ref.\ \cite{Weimann2017EdgeStatesPhotonicCrystal}.

Motivated by the analogy between the effective Hamiltonian \eqref{eq:SSHU2Bloch} and the shape matrix given in Eq.\ \eqref{eq:BlochLiouvillean}, we now formulate the complex Zak phase defined for the master bands of the Bloch Liouvillean.
The NMM bands obtained from a description via LME can be classified topologically in the same fashion as the bands of the Bloch Hamiltonian $H_\text{eff}^{(U_2)}$.
Therefore we define a Zak phase $\nu$ for the NMM bands by \mbox{using} the left- and right-hand eigenstates of the Bloch Liouvillean \eqref{eq:BlochLiouvillean} in Eq.\ \eqref{eq:complexBerry}.

 \section{\label{sec:Comparison}Comparison of both descriptions}

\begin{figure}
	\centering
	\includegraphics{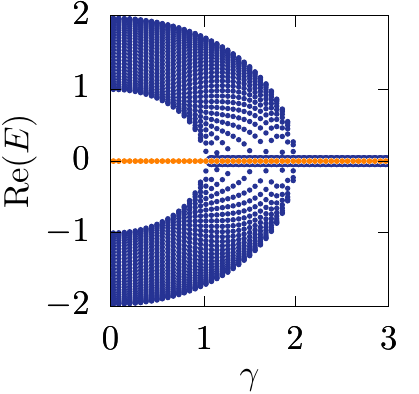}
	\includegraphics{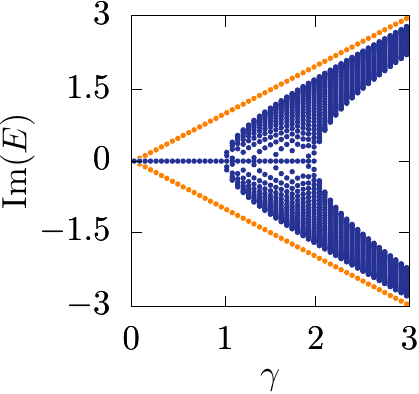}
	\vspace{-4ex}
	\caption{Complex single-particle energy spectrum of the $\mathcal{PT}$-symmetric Hamiltonian $H_\text{eff}^{(U_2)}$ for different dissipative strengths with highlighted differences between trivial ($\theta = 2\gpi/3$, dark blue points) and nontrivial ($\theta = \gpi/3$, additional light orange points) dimerization. Additional features in the nontrivial lattice configuration are caused by zero-energy edge modes.}
	\label{fig:U2ptSpectrum}
\end{figure}

Having outlined the different approaches of modeling dissipation in the SSH model, a key aspect of this work shall be dedicated to a comparison of both methods in order to check their conformity.

While the $\mathcal{PT}$-unbroken regime in the effective theory yields stationary modes, which can also be understood as a non-Hermitian extension of Hermitian quantum mechanics, the interpretation of the $\mathcal{PT}$-broken regime with complex eigenvalues is questionable, 
as the exponential change of the probability density resulting from a time evolution with the effective Hamiltonian may result in an unphysical behavior of the system. 
The applied scheme in this work relies on the analogy between the effective theory and LME outlined in Sec.\ \ref{subsec:effectiveTheory}. 
Starting from the dissipation-free Hamiltonian many-body ground state scenario ($\gamma = 0$), where all single-particle modes with negative and zero energy $\Re(E) \le 0$ (to include edge modes) are occupied, we expect those modes to remain occupied as long as the imaginary part vanishes exactly when dissipation is turned on. 
Whenever a mode breaks $\mathcal{PT}$ symmetry by acquiring a complex eigenvalue, the sign of the imaginary part determines whether the mode is filled up ($+$) or emptied ($-$) in the long-time limit. 
Applying this interpretation we identify a NESS-like many-body state built up from the specified single-particle modes  in the effective framework. 
This state corresponds to the complex many-body energy with a maximum imaginary and minimum real part. 
Thus, we refer to this state as \emph{maximally $\mathcal{PT}$-broken ground state} (MBS), which is unique if all single-particle modes possess non-zero energies.

The further investigation within this section will reveal a good agreement of the MBS and the NESS with respect to the expectation value of the lattice site occupation operators. This is a surprising observation as the construction of both states relies on very different methods. 
The idea behind the MBS is a modification of the many-body ground state due to the effects of dissipation. As we are interested in the ground state the MBS is constructed by a conditional minimization of the energy. Some single-particle modes of the ground state will be affected by the dissipative terms and thus are filled or emptied due to dissipation. 
By contrast the decision whether a master mode is occupied in the NESS is solely determined by the sign of the corresponding eigenvalue's real part. 

Hereinafter we parameterize the tunneling amplitudes accordingly to $t_{1/2}=t (1\mp\Delta\cos(\theta))$.

\begin{figure}
	\centering
	\includegraphics{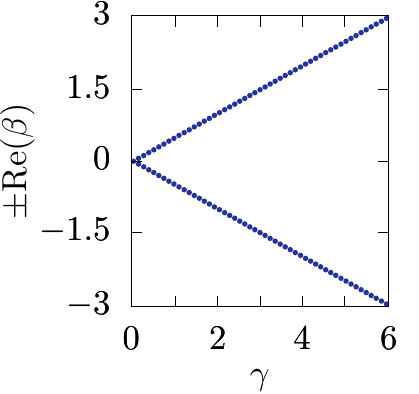}
	\includegraphics{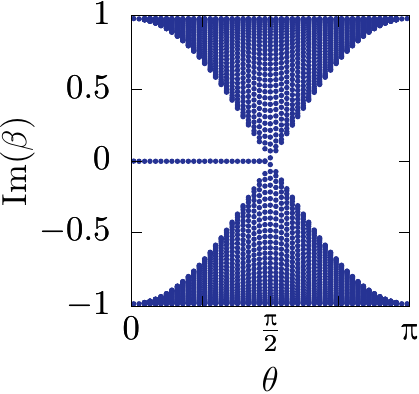}
	\vspace{-4ex}
	\caption{Rapidity spectrum of the Liouvillean shape matrix with dissipative couplings according to $U_2$. The decay rates $\pm \Re(\beta)$ of the NMM shown in the left panel are \emph{independent} of the lattice dimerization (see App.\ \ref{app:proof}). By contrast, the ima\-ginary parts $\Im(\beta)$ depend only on the Hamiltonian and are presented in the right panel for different dimerizations $\theta$. We emphasize that the imaginary rapidity spectrum \emph{exactly} reproduces the energies of the Hermitian SSH model.}
	\label{fig:U2rapiditySpectrum}
\end{figure}

\subsection{Alternating gain and loss}

First we consider an SSH chain ($n = 64$ and \mbox{$t = \Delta = 1$}) subject to the dissipative pattern $U_2$ and compute the complex energy spectrum of the corresponding $\mathcal{PT}$-symmetric Hamiltonian, which was previously discussed in \cite{Weimann2017EdgeStatesPhotonicCrystal, Klett2017PTSymmetry}, as well as the rapidity spectrum of the Liouvillean \eqref{eq:ShapePeriodicBC} (without periodic boundary conditions).

Fig.\ \ref{fig:U2ptSpectrum} shows the complex energies of the system for a varying gain-loss strength, with highlighted differences between the topologically trivial ($t_1 / t_2  = 3$, $\theta=2 \gpi /3$) and nontrivial ($t_1 / t_2 = 1/3$, $\theta=\gpi /3$) dimerization.
We note that the phase transition to the $\mathcal{PT}$-broken regime is in agreement with that of the analytical eigenvalues of the Bloch Hamiltonian given in Eq.\ \eqref{eq:SSHU2Bloch} for the bulk modes (blue points). 
By contrast the nontrivial dimerization possesses $\mathcal{PT}$-broken modes for any non-zero $\gamma$ as the edge states can obviously not be eigenstates of parity-time reflection \cite{Klett2017PTSymmetry} and thus immediately break the symmetry. 
Moreover, the imaginary energy of the edge modes is linear and given by $\pm \ii \gamma$, which follows from the fact that the edge modes are supported only on one of the sublattices $A,B$ \cite{Asboth2016TopologicalInsulators}. 
The real part of the Bloch bands containing the bulk modes bend towards zero for increasing dissipation and eventually each mode breaks the $\mathcal{PT}$ symmetry with a purely imaginary eigenvalue. 
This observation is exactly the expected behavior in the strongly dissipative scenario where the hopping can be considered as a weak perturbation, and in which all lattice sites effectively decouple and yield independent sites being subject to either gain or loss, that is $\lim_{\gamma \to \infty} \Re(E) = 0$ and $\lim_{\gamma \to \infty} \Im(E) = \pm\gamma$.

In Fig.\ \ref{fig:U2rapiditySpectrum} we show the rapidities of the analogue system formulated in the framework of the LME. 
Whereas the presence of a $\mathcal{PT}$-symmetric region in the effective theory suggested a regime with stationary modes despite dissipation, the convergence rates in the Lindblad scenario are equal for each NMM, $\Re(\beta) = \gamma/2$, such that all modes are sensitive to the reservoir regardless of the lattice configuration. 
Interestingly, we find that for the specific pattern with alternating gain and loss the features of the unitary Hamiltonian and the collapse operators decouple, since it can be shown analytically for the periodic system that the rapidities are two-fold degenerate and given by $\beta = \gamma/2 \pm \ii E_{m}/2$, where $E_{m}$ denotes the eigenvalues of the Hermitian SSH model from \mbox{Eq.\ \eqref{equ:sshBlochHamiltonian}} (compare App.\ \ref{app:proof}). 
We will further comment on this remarkable property in the course of this section.

\begin{figure}
	\centering
	\includegraphics{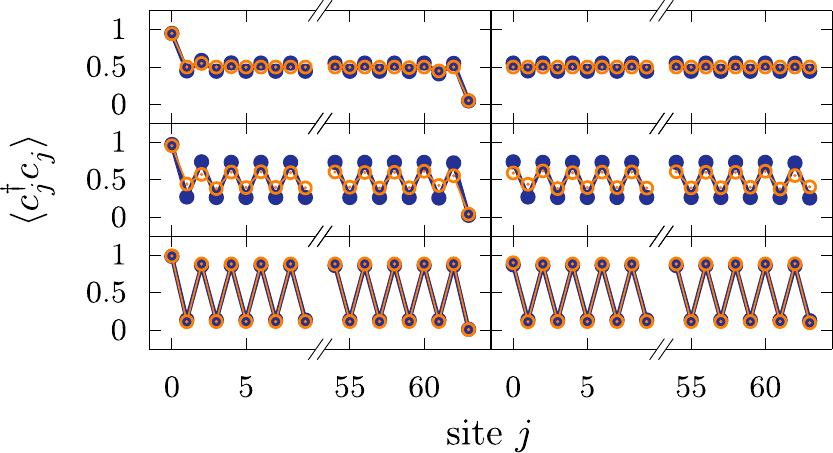}
	\vspace{-1ex}
	\caption{Lattice occupation corresponding to the NESS (blue dots) and MBS (open orange circles) of the SSH model with alternating gain and loss ($\gamma=0.5,1.4,2.5$ from top to bottom). Left panels: Nontrivial dimerization ($\theta = \gpi/3$). Right panels: Trivial dimerization ($\theta = 2 \gpi/3$).}
	\label{fig:U2ss}
\end{figure}

The MBS and NESS lattice occupations derived from the spectra in Figs.\ \ref{fig:U2ptSpectrum}, \ref{fig:U2rapiditySpectrum} are compared in Fig.\ \ref{fig:U2ss}. 
For increasing dissipation, the bulk makes a transition into a progressively staggered configuration, which ultimately leads to entirely filled/empty sites. 
However, the edge modes occurring in the nontrivial lattice dimerization play an important role in both descriptions and are occupied/emptied for finite dissipation. From this it follows that a pronounced occupation at the edges can always be observed whenever the SSH Hamiltonian yields edge modes. 
We note that the MBS qualitatively reproduces the NESS behavior to a good extent, especially the property of half filling. 
Only in the $\mathcal{PT}$-unbroken regime where the effective theory suggests a flat bulk, a slight imbalance between gain and loss sites can already be detected in the Lindblad description (compare top row of Fig.\ \ref{fig:U2ss}). 
In fact, the MBS does not show a staggered bulk in the absence of $\mathcal{PT}$-broken bulk modes (blue bands in Fig.\ \ref{fig:U2ptSpectrum}). 

Carrying on the remarkable agreement of our approaches to model dissipation, we will now address the question of how a topological invariant, which already exists for the effective theory \cite{Garrison1988}, transfers to the Liouvillean.

\subsubsection*{Complex Zak phase}
The question how the notion of topological invariants of Hamiltonian states or bands can be extended to open quantum systems represents a challenge to current research. 

\begin{figure}
	\centering
	\includegraphics{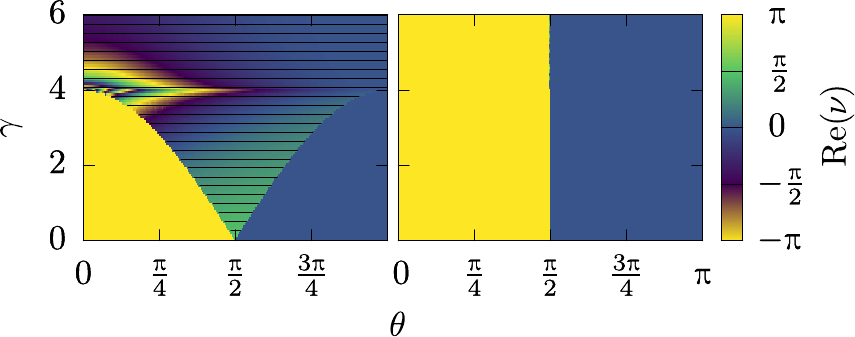}
	\caption{Complex Zak phase $\nu$ of the SSH model ($t=\Delta=1$) with alternating gain and loss. Left panel: Real part of the complex Zak phase obtained from the effective description using a complex $\mathcal{PT}$-symmetric potential. The hatched area marks the $\mathcal{PT}$-broken parameter regime in which the real part of the complex Zak phase is \emph{not} quantized (see App.\ \ref{app:quanizationBerry} for details). Right panel: Zak phase characterizing an NMM band of the Bloch Liouvillean obtained from a description of the system using an LME.}
	\label{fig:phaseU2}
\end{figure}

The complex Bloch bands of the non-Hermitian system can be classified topologically in the $\mathcal{PT}$-unbroken parameter regime by the real part of the complex Zak phase, which is shown in the left panel of Fig.\ \ref{fig:phaseU2}. 
This phase diagram is obtained from a numerical evaluation of Eq.\ \eqref{eq:complexBerry} using the eigenvectors of the non-Hermitian Bloch Hamiltonian given by Eq.\ \eqref{eq:SSHU2Bloch} via the algorithm presented in Ref.\ \cite{we2017}. 
The scenario is in full accordance with the topological band theory of Hermitian systems, and the complex Bloch bands possess a quantized Zak phase, which is well-defined as long as the bands are gapped. 
The system is in a topologically trivial (nontrivial) phase in the $\mathcal{PT}$-unbroken parameter regime for $\theta>\gpi/2$ ($\theta<\gpi/2$). 
In the $\mathcal{PT}$-broken parameter regime the complex Bloch bands are gapless and two exceptional points exist within the Brillouin zone \cite{Weimann2017EdgeStatesPhotonicCrystal}.

Since the quantization of the real part of the complex Zak phase collapses in the $\mathcal{PT}$-broken parameter regime that separates the $\mathcal{PT}$-unbroken regions, a sharp topological phase transition indicated by a discontinuous change of the real part of the Zak phase of $\gpi$ cannot be observed for values of $\gamma>0$.

In the same fashion the NMM bands obtained from a description via an LME can be classified topologically in analogy with the Hermitian band theory.
Therefore we calculate the Zak phase of the NMM bands by evaluating Eq.\ \eqref{eq:complexBerry} \mbox{using} the left- and right-hand eigenstates of the Bloch Liouvillean \eqref{eq:BlochLiouvillean}. 
A numerical evaluation yields the right panel of Fig.\ \ref{fig:phaseU2}. 
For a given dimerization parameter $\theta$ and gain and loss strength $\gamma$ one finds all NMM bands to be characterized by the same quantized phase with a vanishing imaginary part. 
The Zak phase of the NMM bands indicates a trivial (nontrivial) phase for a trivially (nontrivially) dimerized chain, which is in full agreement with the isolated SSH model. 
This observation is found to be inherently linked to the structure of the shape matrix for $U_2$ given in Eq.\ \eqref{eq:BlochLiouvillean}, to which we provide further details in App.\ \ref{app:proof}. 
In fact it turns out that the Liouvillean can be decomposed into outer products. One of the product spaces contains the entire structure of the Bloch Hamiltonian, such that the Berry phase of the transport along the Brillouin zone is inherited from the Hermitian case. 
Thus, the NMM bands possess a Zak phase of $\gpi$ if the Hermitian SSH chain is dimerized topologically non-trivial ($\theta<\gpi/2$), and a Zak phase of $0$ in case of trivial dimerization ($\theta>\gpi/2$). 
For $\theta=\gpi/2$ the NMM bands touch each other and the Zak phase is not well-defined.
In contrast to the complex Zak phase of the effective $\mathcal{PT}$-symmetric description this quantization does not require a special structure of the eigenvalue spectrum of the Bloch Liouvillean, the quantization is rather ensured by the chiral symmetry of the Hermitian SSH model from which the Zak phase is inherited to the eigenstates of the Bloch Liouvillean (see App.\ \ref{app:proof}).

Hence, in the case of alternating gain and loss we can define a dissipative analogue of the Zak phase for the NMM bands, which carries the information about the topology of the Hamiltonian and is not affected by the strength of dissipation. 
This Zak phase of the NMM bands is in perfect agreement with the complex Zak phase obtained from the effective approach in the para\-meter regime where the complex energy spectrum of the non-Hermitian Hamiltonian is entirely real-valued ($\mathcal{PT}$-unbroken phase).
We comment on the stability of the complex Zak phase for the disordered case in App.\ \ref{App:disorder}. 

 \subsection{Gain and loss at the boundaries}

In the following we compare the interpretation of the effective theory introduced in the beginning of this section with the LME description of an SSH chain ($n = 64$ and $t = \Delta = 1$) subject to a reservoir of the type $U_1$ (cf.\ Fig.\ \ref{fig:model} and Eq.\ \eqref{eq:PotU1} resp.\ Eq.\ \eqref{eq:LindbladOpsU1}).

\begin{figure}
	\centering
	\includegraphics{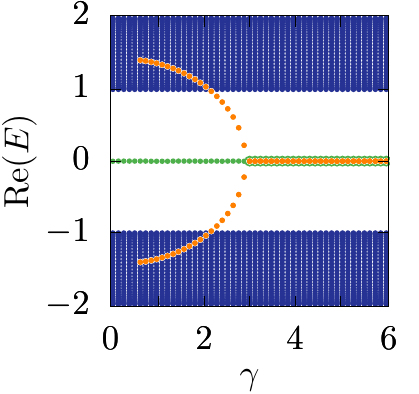}
	\includegraphics{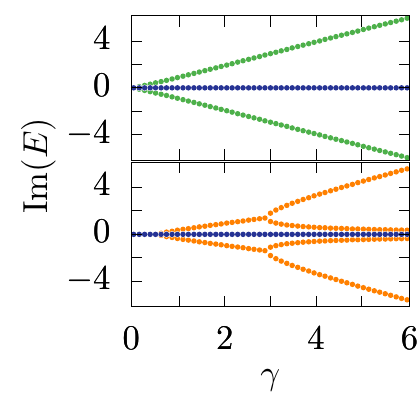}
	\vspace{-4ex}
	\caption{Complex single-particle eigenvalue spectrum of the $\mathcal{PT}$-symmetric Hamiltonian of the SSH model subject to the complex potential $U_1$ with $\theta = \gpi/3$ (blue and green points in the left panel, and right top) respectively $\theta = 2\gpi/3$ (blue and orange points in the left panel, and right bottom). The colors of the imaginary part branches on the right correspond to the real part of the complex energies shown on the left.}
	\label{fig:U1ptSpectrum}
\end{figure}

Fig.\ \ref{fig:U1ptSpectrum} shows the single-particle spectrum of the non-Hermitian Bloch Hamiltonian in dependence of the dissipative strength $\gamma$ for a topologically trivial ($t_1 / t_2  = 3$, $\theta=2 \gpi /3$) and nontrivial ($t_1 / t_2 = 1/3$, $\theta=\gpi /3$) dimerization. In the latter, the edge states are again found to be $\mathcal{PT}$-broken for all values $\gamma>0$. 
The imaginary part of the complex energy is restricted by $|\t{Im}(E)|<\gamma$ and the localisation of the edge modes increases with $\gamma$, such that one finds $\lim_{\gamma \to \infty} \Im(E) = \pm\gamma$ for their energy. 

The same holds in the strongly dissipative scenario of the trivial SSH chain. 
After the $\mathcal{PT}$ phase transition the real part of the complex energies of the modes located mainly at the boundary unit cells bend towards zero (orange points in Fig.\ \ref{fig:U1ptSpectrum}). 
These modes decouple while going through a bifurcation at $\gamma=3$ and collapse into (i) modes hosted solely by the dissipative site at the boundary and (ii) edge modes in the nontrivially dimerized Hermitian subsystem in the limit $\gamma \rightarrow \infty$.
While the latter corres\-pond to the eigenvalue branches whose complex energy converges towards zero, the eigenvalues of the others are given by $\pm \ii \gamma$ in the strongly dissipative limit. The effect of the modes with approximately zero total energies for large values of $\gamma$ becomes visible in the MBS, which is discussed later, and illustrates that these states indeed correspond to internal edge states of the subsystem, i.e.\ the SSH chain without the dissipative sites.

\begin{figure}
	\centering
	\includegraphics{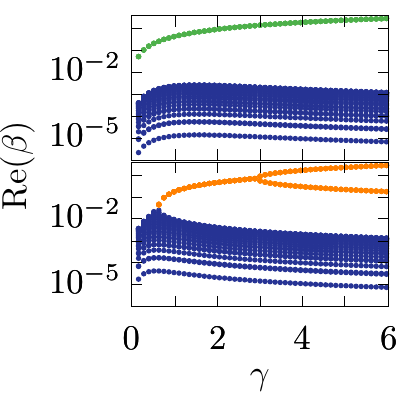}
	\includegraphics{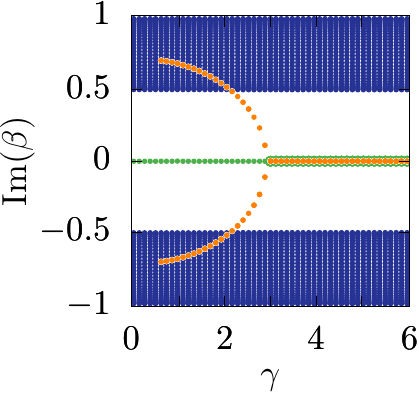}
	\vspace{-4ex}
	\caption{Rapidity spectrum of the Liouvillean of the SSH model with gain and loss at the edges (Lindblad operators $L_{\mu}^{(U_1)}$) with $\theta = \gpi/3$ (left top, and blue and green points in the right panel) respectively $\theta = 2\gpi/3$ (left bottom, and blue and orange points in the right panel). The colors of the real part branches correspond to the imaginary part of the rapidities shown on the right.}
	\label{fig:U1rapiditySpectrum}
\end{figure}

Interestingly, one finds considerable resemblances between the complex energies and the rapidities following from a description via LME. 
The imaginary part of the rapidity spectrum (cf.\ Fig.\ \ref{fig:U1rapiditySpectrum}) reproduces the real part of the complex energy spectrum (cf.\ Fig.\ \ref{fig:U1ptSpectrum}).
Also the real part of the rapidities (cf.\ left panel in Fig.\ \ref{fig:U1rapiditySpectrum}) and the imaginary part of the single-particle modes ($\t{Im}(E)$ cf.\ right side in Fig. \ref{fig:U1ptSpectrum}) show significant \mbox{similarities. Only the} exactly vanishing imaginary parts leading to stationary bulk modes in the effective description differ from their counterpart in the LME framework where bulk NMMs are always characterized by a small (but nonzero) relaxation rate that guarantees the uniqueness of the NESS.  
Furthermore, the bifurcations in the rapidity spectrum and the complex energy spectrum approximately occur at the same parameter values. Generally, only modes hosted by sites coupled to the reservoir are strongly affected by dissipation, which also causes a change of the lattice site occupation. This is further investigated by comparing the NESS and the MBS.

\begin{figure}[b]
	\centering
	\includegraphics{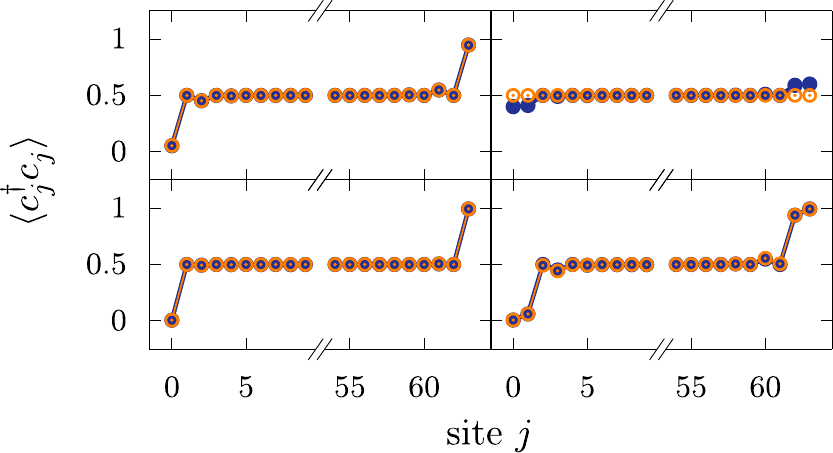}
	\vspace{-1ex}
	\caption{Comparison of the lattice site occupation profile of the NESS (blue circles) and the corresponding MBS (open orange circles) for $t=\Delta=1$ and $\theta = \gpi/3$ on the left respectively $\theta = 2 \gpi/3$ on the right and $\gamma=0.25,4$ from top to bottom.}
	\label{fig:U1ss}
\end{figure}

We show MBS and NESS lattice occupations for a reservoir of the type $U_1$ in Fig.\ \ref{fig:U1ss}. 
Due to its construction the MBS contains the single-particle edge modes which are occupied due to the  effect of the gain and loss of particles.
For increasing dissipative strengths the edge mode localization at the lattice boundaries becomes more pronounced and the mode extends less into the bulk.

In case of a trivial chain and values of $\gamma \lesssim 0.5$ the $\mathcal{PT}$-symmetric Hamiltonian is $\mathcal{PT}$-unbroken (see Fig.\ \ref{fig:U1ptSpectrum} right bottom) and the corresponding MBS, which has a real-valued energy is equivalent to a Mott state at half filling (cf.\ Fig.\ \ref{fig:U1ss} right top) \cite{Muth2008HalfIntegerMI}. 
In this parameter regime the MBS does not reproduce the NESS lattice occupations. 
However, the trivial chain yields lattice occupation profiles of the NESS and the MBS, which match perfectly for large values of $\gamma$.
In this case the sites at the edges of the lattice act as a connection between the reservoir and the rest of the system. 
For large values of $\gamma$ one can assign the edge sites (the occupations of which are approximately fixed to 0 resp.\ 1) to the reservoir and finds the NESS (MBS) in the subsystem to reproduce the NESS (MBS) of the non-trivial phase (cf.\ bottom of Fig.\ \ref{fig:U1ss}). 
This effect is caused by the inversion of the sublattice's dimerization due to the cut-off of dissipative sites leaving a non-trivially dimerized system. 

We note that this decoupling of dissipative sites from their surrounding was already observed in the strongly dissipative scenario of alternating gain and loss, where the lattice occupation is found to be completely staggered (cf.\ bottom panel of Fig.\ \ref{fig:U2ss}).

 \section{Conclusion}
One of the main efforts of this work was a comparison between two approaches for describing dissipative quantum systems, viz.\ Lindblad master equations and an effective theory using $\mathcal{PT}$-symmetric on-site potentials. 
For alternating gain and loss we have defined topological invariants which characterize the complex Bloch bands in the $\mathcal{PT}$-unbroken regime and the corresponding master bands of the Liouvillean to match perfectly.
The topo\-logical invariants rely on the idea that both frameworks allow for a generalization of the Hermitian Zak phase to the dissipative scenario where the entire information is contained in a non-Hermitian Bloch Hamiltonian resp.\ Bloch Liouvillean.
In particular we showed that the quantization of the complex Zak phase's real part is protected by the $\mathcal{PT}$ symmetry in the $\mathcal{PT}$-unbroken parameter regime.
By contrast the spectral properties of the Liouvillean yield a decoupled structure that contains the bands of the Hamiltonian as a subset. Hence, the quantization of the Zak phase of the master bands can be traced back to the topological properties of the Hermitian SSH Hamiltonian.
An investigation of the effect of hopping disorder which respects the symmetry properties of the system shows that the complex energy spectra of the system do neither undergo a $\mathcal{PT}$ phase transition nor a gap closing, such that complex the Zak phase is robust against a sufficiently small disorder. The same holds for the Zak phase of the master bands as it is inherited from the Hermitian SSH model which is robust against hopping disorder. 

Working with the effective theory we have introduced and justified an interpretation to obtain the long-term fixed-point of the system, by regarding imaginary parts of the energies in the $\mathcal{PT}$-broken regime as decay rates.
The resulting lattice occupation shows a remarkable resemblance to the non-equilibrium steady state extracted from the Liouvillean. 
In the regime of weak dissipation where the bulk modes do not break the $\mathcal{PT}$ symmetry in the effective description, the steady state of the effective description (MBS) and the non-equilibrium steady state (NESS) of the Lindblad master equation disagree as the expectation values of the occupation operator of the steady state in the effective framework is not influenced by the reservoir in the $\mathcal{PT}$-unbroken parameter regime.

Using a reservoir coupled to the edges of the SSH chain we have found that the edge modes are occupied or emptied with progressing time evolution of the system. 
The dissipative sites effectively decouple from their surrounding in the strongly dissipative regime in such a way that by turning on dissipation in the trivial regime the steady state of the subsystem (where the two edge sites are ranked to belong to the reservoir) reproduces the steady state of the nontrivial SSH chain.

\appendix
\section{\label{app:quanizationBerry}Quantization of the geometric phase}

As shown in the work of Hatsugai \cite{Hatsugai2006}, the Berry phase $\nu$ of closed systems (described by a Hermitian Hamiltonian) is quantized in the presence of an antiunitary symmetry ($\mathcal{PT}$ in the context of this work) and takes integer multiples of $\pi$. The argument carries over to non-Hermitian systems described by a $\mathcal{PT}$-symmetric Hamiltonian $H(\bm{\alpha})$ parameterized by $\bm{\alpha}\in \mathbb{P}$ in parameter space $\mathbb{P}$:
For a dual pair of biorthogonal eigenstates $(\bra{\chi_n(\vec{\alpha})}, \ket{\phi_n(\vec{\alpha})})$ of $H$ with a non-degenerate eigenvalue $\lambda_n(\vec{\alpha})$, that is $H \ket{\phi_n} = \lambda_n \ket{\phi_n}$ and $\bra{\chi_n} H = \lambda_n \bra{\chi_n}$,
one finds the real part of the complex Berry phase $\nu_n$ \cite{Garrison1988}, which is picked up by the eigenstates during the transport along a loop $\mathcal{C} \subset \mathbb{P}$ to be quantized in integer multiples of $\pi$ if $(\bra{\chi_n}, \ket{\phi_n})$ are eigenstates of $\mathcal{PT}$ (which implies $\lambda_n \in \mathbb{R}$). 

To see this, we consider the Berry phase of $(\bra{\chi_n}, \ket{\phi_n})$ obtained by integration of the complex Berry connection $\vec{\mathcal{A}}_n^{}(\vec{\alpha}) = \ii \braket{\chi_n^{} | \vec{\nabla_\alpha} | \phi_n^{}}$ along $\mathcal{C}$,
\begin{equation} 
	\label{eq:nonHermBerry}
	\nu_n= \ii \oint_{\mathcal{C}} \braket{\chi_n
	| \vec{\nabla_{\alpha}} | \phi_n}
	\cdot \dd\vec{\alpha}.
\end{equation} 
Expanding the states in a fixed Cartesian basis $\{ \ket{e_j}, j = 1, \dots, \dim(H) \,|\, \braket{e_j | e_k} = \delta_{jk} \}$ by 
$\bra{\chi_n^{}}=\sum_{j} \chi_{nj^{}}(\vec{\alpha}) \bra{e_j^{}}$ and $\ket{\phi_n^{}}= \sum_{j} \phi_{nj^{}}(\vec{\alpha}) \ket{e_j^{}}$, the $k$th component of the Berry connection reads $\vec{\mathcal{A}}_{n,k} = \ii \sum_{j} \chi_{nj}^{} \, \partial_{\alpha_k} \phi_{nj}^{}$.

The latter can be related to the Berry connection $\vec{\mathcal{A}}_n^{\mathcal{PT}}$ of the $\mathcal{PT}$-symmetric partners (with the abbreviation $\bra{x}\mathcal{PT} \equiv \bra{x^{\mathcal{PT}}}$, $\mathcal{PT} \ket{x} \equiv \bra{x^{\mathcal{PT}}}$) by evaluating the action of $\mathcal{PT}$ as $\bra{\chi_n^{\mathcal{PT}}}=\sum_{j}^{} \chi_{nj}^{\ast} \bra{e_j^{\mathcal{PT}}}$ and $\ket{\phi_n^{\mathcal{PT}}}=\sum_{j}^{} \phi_{nj}^{\ast} \ket{e_j^{\mathcal{PT}}}$. A straightforward calculation leads to the important relation $\vec{\mathcal{A}}_n^{\mathcal{PT}} = -\vec{\mathcal{A}}_n^{\ast}$, which directly implies
\begin{equation}\label{eq:BerryPhaseRelationI}
\nu_n^{} =-\oint_{\mathcal{C}} \vec{\mathcal{A}}_n^{\mathcal{PT}\ast} \cdot \dd\vec{\alpha} = -\nu_n^{\mathcal{PT}\ast}.
\end{equation}

A second important relation follows from the fact that the action of $\mathcal{PT}$ onto an eigenstate of $H$ is nothing but a $U(1)$ gauge transformation if the eigenvalue is non-degenerate and real. Since the $\mathcal{PT}$-symmetric Hamiltonian satisfies the commutation relation $[H,\mathcal{PT}]=0$, the application of $\mathcal{PT}$ on the (right) eigenvalue equation yields
$ H \ket{\phi_n^{\mathcal{PT}}}  =  \lambda_n^{\ast} \ket{\phi_n^{\mathcal{PT}}}$,
which reduces to the eigenvalue relation of $\ket{\phi_n}$ in case of a real eigenvalue $\lambda_n \in \mathbb{R}$, that is if $\ket{\phi_n^{}}$ (or $\bra{\chi_n^{}}$) are right (or left) eigenstates of $\mathcal{PT}$ \cite{Bender2007}. In combination with the assumption of a non-degenerate eigenvalue $\lambda_n$ and the normalization convention $\braket{\chi_n^{\mathcal{PT}} | \phi_n^{\mathcal{PT}}} = 1$, this implies that $\mathcal{PT}$ corresponds to a $U(1)$ phase transformation
$\ket{\phi_n^{\mathcal{PT}}} = \ee^{\ii \varphi_n^{}} \ket{\phi_n^{}}$ and $\bra{\chi_n^{\mathcal{PT}}} = \ee^{-\ii \varphi_n^{}} \bra{\chi_n^{}}$. Given that case, the Berry connection of the $\mathcal{PT}$ partners evaluates to 
$\vec{\mathcal{A}}_n^{\mathcal{PT}} =  \vec{\mathcal{A}}_n - \vec{\nabla_\alpha} \varphi_n$.

Using the fact that the basis states have to be single-valued in parameter space such that one can find an integer $z\in \mathbb{Z}$ for which $\oint_{\mathcal{C}} \vec{\nabla_\alpha} \varphi_n^{} \cdot \dd\vec{\alpha}  = 2\pi z$ holds, the latter relation can be integrated along $\mathcal{C}$ leaving
\begin{equation}\label{eq:BerryPhaseRelationII}
	\nu_n^{\mathcal{PT}} = \oint_{\mathcal{C}} ( \vec{\mathcal{A}}_n^{} - \vec{\nabla_\alpha}\varphi_n^{}) \cdot \dd\vec{\alpha} = \nu_n^{} -2\pi z.
\end{equation} 

Finally, the combination of Eqs.\ \eqref{eq:BerryPhaseRelationI} and \eqref{eq:BerryPhaseRelationII} induce the quantization of the real part of the complex Berry phase in the $\mathcal{PT}$-unbroken regime,
\begin{equation}
	\label{eq:finalQuantisation}
	\nu_n^{} = -\nu_n^{\mathcal{PT}\ast} = - (\nu_n^{} - 2\pi z )^{\ast}
	\implies
	\Re(\nu_n^{}) = \pi z. 
\end{equation} 
Hence, in this case, the real part of the complex Berry phase can be constrained to a strict quantization protected by $\mathcal{PT}$ symmetry.

\section{\label{app:proof}Rapidities of the dissipative pattern \texorpdfstring{\bm{$U_2$}}{U2}}

We give the proof that the imaginary part of the rapidities of the SSH model subject to a dissipative pattern described by $U_2$ in the context of an LME reproduces the Hamiltonian spectrum, while the real part is solely determined by the gain-loss strength. 
To do so, we consider the eigenvalue equation $\bm{A} \bm{v}_j = \beta_j \bm{v}_j$ of the $4n\times 4n$ dimensional anti-symmetric ($\bm{A}^T = -\bm{A}$ in the basis of the Hermitian Majorana maps $\hat{a}_i$, compare Eq.\ \eqref{eq:ShapePeriodicBC}) shape matrix in order to determine the $j$th \emph{normal master mode (NMM)} $\bm{v}_j$ with an associated rapidity $\beta_j$. 
In the scenario with alternating gain and loss, a basis transformation accounted for by a Fourier transform leads to the Bloch form of $\bm{A}$ in Eq.\ \eqref{eq:BlochLiouvillean}, which we repeat here for convenience,
\begin{align*} 
	\bm{A} =  \frac{1}{2} \sum_k \ket{k}\!\bra{k} \otimes& \begin{pmatrix}
	\gamma \bm{\Gamma}_\text{g} & -(t_1 + \mathrm{e}^{\ii k} t_2) \bm{T}
	\\
	-(t_1 + \mathrm{e}^{-\ii k} t_2) \bm{T} & \gamma \bm{\Gamma}_\text{l} 
	\end{pmatrix},
\end{align*}
where the matrices $\bm{\Gamma}_\text{g(l)} = - \identity_{2} \otimes \sigma_y \varpm \sigma_y \otimes \left( \ii \sigma_x  + \sigma_z\right)$ and  $\bm{T} = -\ii\sigma_y \otimes \identity_{2}$ can be considered as a generalization of the non-Hermitian effective Hamiltonian scenario. 
In order to see how eigenvectors of this equation can be constructed, the $8 \times 8$ Bloch Liouvillean (cf.\ Eq.\ \eqref{eq:BlochLiouvillean}) is further decomposed into a sum of three Kronecker products of $2\times2$-dimensional matrices as follows,
\begin{align}
	\label{eq:BlochLiouvilleanDecomposed}
	\begin{split}
		\bm{A} = \frac{1}{2} \sum_k \ket{k}\!\bra{k} \otimes  \Big[ -\gamma  \identity_{2} &\otimes \identity_{2} \otimes \sigma_y
		\\[-2ex]
		 +\gamma \sigma_z &\otimes \sigma_y \otimes(\ii \sigma_x +\sigma_z) 
		\\
		 -\ii H_{\text{B}} &\otimes \sigma_y \otimes \identity_{2}\Big],	
	\end{split}
\end{align}	
where the $2\times2$ Bloch Hamiltonian $H_\text{B}$ of the Hermitian SSH model (compare Eq.\ \eqref{equ:sshBlochHamiltonian}) naturally appears. 
Starting off with this form, we attempt to construct master modes $\bm{v}_j$ which are eigenvectors of \emph{each} of the three summands in Eq.\ \eqref{eq:BlochLiouvilleanDecomposed}. 
In doing so, one notices for the square bracket term that the operators of the first product space commute, that is $[\sigma_z, H_\text{B}] = 0$ (and trivially $[\sigma_z, \identity_2] = [H_\text{B}, \identity_2] = 0$) and thus common eigenvectors of all three summands, restricted to the first product space, exist. 
The same statement holds for the second product space, but is violated by the third one, as $[\sigma_y, \ii \sigma_x + \sigma_z] \neq 0$.
However, the matrix $\ii \sigma_x + \sigma_z$ has an eigenvector $(1, \ii)^T$ with zero eigenvalue, which is simultaneously an eigenvector of $\sigma_y$ with eigenvalue $1$.

This having been said, a procedure to construct $2n$ of the overall $4n$ eigenvalues and eigenvectors of the shape matrix can be formulated: 
The NMM $\bm{v}_j$ is constructed by a Kronecker product of the momentum space component and the three internal spaces, where the vector in the last internal space is always given by $(1, \ii)^T$, and consequently the second term in Eq.\ \eqref{eq:BlochLiouvilleanDecomposed} does not contribute to the eigenvalue equation $\bm{A} \bm{v}_j = \pm\beta_j \bm{v}_j$. 
For the remaining two summands and internal spaces, $\-\gamma \identity_2 \otimes \identity_2 - \ii H_\text{B} \otimes \sigma_y$, it was argued above that the constituents' possess a common set of eigenvectors. 
As the eigenvalues of $H_\text{B}$ are given by the energies $E_m(k) = \pm \sqrt{t_1^2 + t_2^2 + 2 t_1 t_2 \cos(k)}$ of the Hermitian SSH model and those of $\sigma_y$ are $\pm 1$, the two internal product spaces give rise to the $2n$ combinations $-\gamma - \ii E_m(k) \cdot (\pm 1)$, which results in the rapidities
\begin{align}
	-\beta_j = -\frac{1}{2} \big( \gamma \pm \ii E_m(k) \big)
\end{align}
with $k = 0, 2\gpi/(n/2), \dots, 2\gpi(n/2 - 1)/(n/2)$ and \mbox{$m = 1, 2$},
such that each value appears twice, leading to $n/2$ two-fold degenerate rapidities. 
We explicitly emphasize that the decay rate $\Re(\beta_j)$ of an NMM $\bm{v}_j$ is solely determined by the gain-loss strength $\gamma$, while the imaginary part reproduces the energy spectrum of the SSH model. 

For completeness, the remaining $2n$ eigenvalues of $\bm{A}$ can be obtained by making use the antisymmetry of $\bm{A}$, which directly implies the existence of the remaining eigenvalues $1/2 ( \gamma \pm \ii E_m(k))$.

To conclude, we note that the factorization of the Bloch Liouvillean given in Eq.\ \eqref{eq:BlochLiouvilleanDecomposed} already shows that in the case of alternating gain and loss the Zak phase of the NMM bands is quantized and complies with the Zak phase of the Hermitian SSH model.
Therefore we point to the first product space which contains the Bloch Hamiltonian $H_\text{B}$ of the Hermitian SSH model (compare Eq.\ \eqref{equ:sshBlochHamiltonian}). All other operators that occur in the factorization do not depend on $k$ and therefore only the component of the eigenstate of the Bloch Liouvillian belonging to the first product space will acquire a phase when the system is transported through the Brillouin zone. This phase is exactly the Zak phase of the Hermitian SSH model as the corresponding eigenstate is precisely the one of the Hermitian Bloch Hamiltonian $H_\text{B}$ that describes the Hermitian SSH model.

\section{Robustness to disorder}
\label{App:disorder}
We now investigate the effect of disorder that respects the symmetries of the system ($\mathcal{PT}$ and $\Lambda$ cf.\ Eq.\ \eqref{eq:SymmetryProperty}) on the complex energy and the rapidity spectra.
To do so, the tunneling amplitudes of each unit cell $j = 1, \dots, n/2$ are randomly exposed to disorder, described by a disorder strength $R$, of the type
\begin{align} \label{eq:disorder}
	\begin{split}
		t_1 &\to \tilde{t}_1 = t_1 + R \xi_j \left| t_1 - t_2  \right|,
		\\
		t_2 &\to \tilde{t}_2 = t_2 - R \xi_j \left| t_1 - t_2 \right|,
	\end{split}
\end{align}
retaining the dimerization on average. The random variables $\xi_j \in (-1; 1)$ are chosen from a uniform distribution in a symmetric manner such that $\xi_j = \xi_{n / 2 + 1 - j}$ .

\begin{figure}
	\centering
	\includegraphics{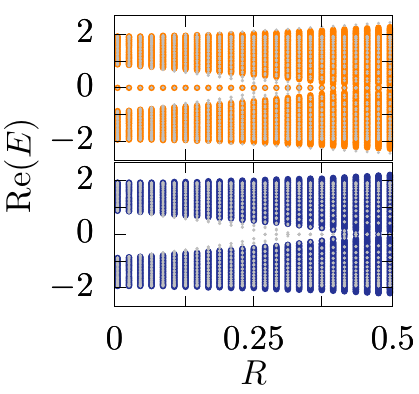}
	\includegraphics{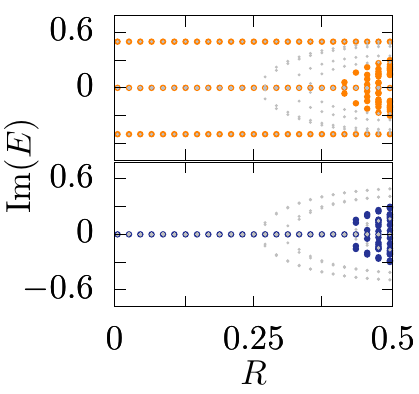}
	\vspace{-4ex}
	\caption{Complex energy spectrum of the $\mathcal{PT}$-symmetric Hamiltonian $H_{\t{eff}}^{(U_{2})}$ with 64 sites subject to disorder of the form \eqref{eq:disorder} for 100 random realizations at $\gamma=0.5$ for $\theta =\gpi/3, 2\gpi/3$ in the top resp.\ bottom row. Gray dots indicate the extreme scenario where $\xi_m=1$ for all $m$.}
	\label{fig:ptDisorderU2}
\end{figure}

Considering the scenario of alternating gain and loss this form of disorder can spoil the topological invariant in the following two ways.

(i) Inversion of the dimerization: For the case of the SSH model ($\gamma = 0$), the critical value of the disorder strength $R_{\text{c}_{1}}$ by which an arbitrary dimerization could in principle be deformed to the homogeneous hopping case $\tilde{t}_1 = \tilde{t}_2$ where the energy gap closes is found to be $R_{\text{c}_{1}} = 0.5$.

(ii) $\mathcal{PT}$ phase transition in the effective theory: Considering the scenario of alternating gain and loss where a unit cell is $\mathcal{PT}$-unbroken as long as $\left| \tilde{t}_1 - \tilde{t}_2 \right| > \gamma$ one finds that even in the worst case scenario $\left| \tilde{t}_1 - \tilde{t}_2 \right| > (1 - 2R) \left|t_1 - t_2\right|$ and thus $R_{\text{c}_{2}} = 1/2 ( 1 - \gamma / (t_1 - t_2))$ \cite{Weimann2017EdgeStatesPhotonicCrystal}.
In the same manner a randomized variation of the on-site potentials can affect the point of the $\mathcal{PT}$ phase transition.

\begin{figure}
	\centering
	\includegraphics{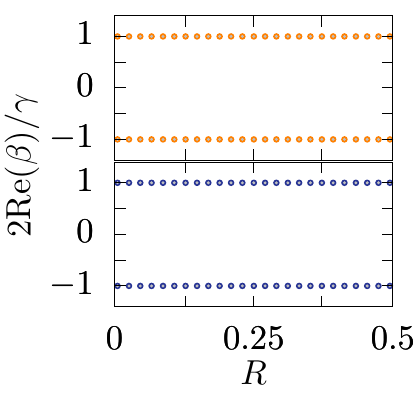}
	\includegraphics{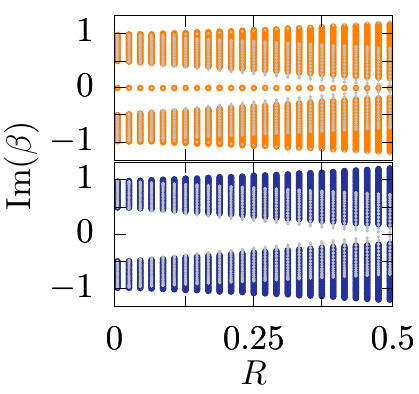}
	\vspace{-4ex}
	\caption{Rapidity spectrum of the Liouvillean describing the dissipative SSH model with 64 sites with alternating gain and loss subject to disorder of the form \eqref{eq:disorder} for 100 random realizations for $\theta =\gpi/3, 2\gpi/3$ in the top resp.\ bottom row. Note the shown spectrum does not depend on $\gamma$. Gray dots indicate the extreme scenario where $\xi_m=1$ for all $m$.}
	\label{fig:LindbladDisorderU2}
\end{figure}

In the following we illustrate both cases using an exemplary choice of parameters. Fig.\ \ref{fig:ptDisorderU2} shows the complex eigenvalue spectrum of the effective Hamiltonian $H_{\t{eff}}^{(U_{2})}$ with alternating gain and loss at $\gamma=0.5$. For both the nontrivial and trivial dimerization one expects the $\mathcal{PT}$ phase transition at $R_{\text{c}_{2}}=0.25$. However, due to the finite number of possible disorders the transitions are shifted to larger values of $R$. The extreme scenario where $\xi_m=1$ for all $m$ is also show for completeness and yields the expected transition point $R_{\text{c}_{2}}=0.25$. For smaller values of $R$ the periodic system is $\mathcal{PT}$-unbroken and possesses two energetically separated bands.  
We verified this property for a variety of parameter sets.

Next we apply the disorder pattern \eqref{eq:disorder} to the Liouvillean of the SSH model coupled to a reservoir describing alternating gain and loss. 
From App.\ \ref{app:proof} we know the structure of the rapidity spectrum without disorder. The real part is determined by the strength of the dissipative coupling whereas the imaginary part of the rapidity spectrum is given by the scaled energy spectrum of the Hermitian SSH model. Therefore we expect that a randomized variation of the tunneling amplitudes only modifies the imaginary part of the rapidities in such a way that for an infinite number of random numbers a gap closing occurs at $R_{\text{c}_{1}} = 0.5$.
Fig. \ref{fig:LindbladDisorderU2} shows the scaled rapidity spectrum of the SSH chain subject to alternating gain and loss. For both the nontrivial and trivial dimerization the finite number of used sets of random numbers causes the gap in the imaginary part to be still present for $R = 0.5$. Again the extreme scenario with $\xi_m=1$ for all $m$ yields the expected gap closing at $R_{\text{c}_{1}} = 0.5$. 
Hence the quantization of the Zak phase picked up by the master modes is not spoiled by disorders with $R<R_{\text{c}_{2}}$ since it relies on the topological invariance of the Hermitian SSH Bloch Hamiltonian as illustrated in App.\ \ref{app:proof}.

\end{document}